\documentclass[12pt, a4paper]{article}
\usepackage[left=3cm,right=3cm,top=3cm, bottom=2.5cm]{geometry}
\usepackage{color}
\usepackage{rotating}
\usepackage{setspace}
\usepackage[super]{natbib}
\usepackage{amssymb}
\usepackage{amsmath}
\usepackage{graphicx}
\usepackage{titlesec}
\usepackage{setspace}
\usepackage[short]{datetime}
\usepackage{comment}
\usepackage[markers,nolists]{endfloat}

\newcommand{\noi}{\noindent}

\begin{document}

\title{Benchmarking survival outcomes:\\A funnel plot for survival data}
\author{Hein Putter, Dirk-Jan Eikema, Liesbeth C.~de Wreede, Eoin McGrath,\\Isabel S\'{a}nchez-Ortega, Riccardo Saccardi, John A.~Snowden, Erik W.~van Zwet}
\date{\today}
\maketitle{}

\begin{abstract}
  Benchmarking is commonly used in many healthcare settings to monitor clinical performance, with the aim of increasing cost-effectiveness and safe care of patients. The funnel plot is a popular tool in visualizing the performance of a healthcare center in relation to other centers and to a target, taking into account statistical uncertainty. In this paper we develop methodology for constructing funnel plots for survival data. The method takes into account censoring and can deal with differences in censoring distributions across centers. Practical issues in implementing the methodology are discussed, particularly in the setting of benchmarking clinical outcomes for hematopoietic stem cell transplantation. A simulation study is performed to assess the performance of the funnel plots under several scenarios. Our methodology is illustrated using data from the EBMT benchmarking project.\\[1cm]
\end{abstract}

\noi
\onehalfspacing

\section{Introduction}
\label{sec:intro}

Benchmarking has become mandatory in many healthcare settings for complex procedures and is used by competent authorities, regulators, payers and patients to monitor clinical performance, with the aim of increasing cost-effectiveness and safe care of patients. Providing a mechanism that allows an easily understandable and fair comparison between healthcare centers is important, because stakes are high, both for future patient well-being and possibly for future funding of the centers. Benchmarking is a challenging task, because of data collection issues and differences between centers in case mix and socio-economic factors that have to be adequately accounted for. In statistical terms, it is important to separate random and systematic differences between centers in a way that is accessible to non-statisticians. The funnel plot, introduced for this purpose by Spiegelhalter~\cite{Spiegelhalter05}, has been generally accepted by healthcare quality researchers as providing an interpretable graphical summary of a collection of centers' relative performance. To date, no extension of the funnel plot to allow for time-to-event data has been proposed.

The current methodological work is the result of an international collaboration aiming at comparing one-year mortality for centers performing hematopoietic stem cell transplantation (HSCT) within the EBMT (www.ebmt.org). HSCT is a treatment associated with high mortality (especially allogeneic HSCT), due to lack of disease control, and effects of transplantation and pre-treatment. It is used for a wide range of diseases, mainly haematologic malignancies, with a large diversity of patient selection across different countries and with many aspects of pre-treatment choices to be considered. Because of the relative rarity of HSCT there is a need for international collaboration. This has been the reason for the establishment of a registry within the EBMT, in which data from a large proportion of HSCT's from practically all countries in Europe and some outside have been registered in a uniform way. In several countries inside and outside Europe, benchmarking systems have already been established for HSCT, but access to these systems has been limited to those countries, and methodology varied across countries~\citep{snowden2020benchmarking}. The EBMT and Joint Accreditation Committee of ISCT and EBMT (JACIE) have therefore established a Clinical Outcomes Group to develop and introduce a universal system accessible across EBMT members. In this paper we report on the methodology underlying this international registry-based risk-adapted benchmarking system for HSCT survival outcomes across the diverse health services and cultures within EBMT.

Our aim is to go beyond in-hospital death or 30-days mortality, because the whole first year post-transplant is associated with considerably high mortality rates. Since follow-up is far from complete even in the first year post-transplant, the methodology has to take account of censoring, which may well differ substantially across centers. More specifically, the objective of this paper to extend the funnel plot to survival outcomes.

\subsection{Benchmarking and funnel plots}

The aim is to compare the performance of individual centers to some benchmark. Most commonly this benchmark concerns some binary (yes/no) outcome or ``indicator'', such as the occurrence of  a particular type of complication, or in-hospital death. In this subsection we stick to this binary case. The benchmark can be either absolute, set by an external target, or relative, determined by some overall average. Funnel plots were proposed by \cite{Spiegelhalter05} as an alternative to league tables. They have now become the standard approach to evaluating quality of care~\citep{kunadian2008cumulative, mayer2009funnel, jacobs2011variation}. To construct a funnel plot we need:
\begin{enumerate}\setlength{\itemsep}{-4pt}
\item An {\em observed} number $O$, for example, the observed number of complications in one year at a particular center;
\item An {\em expected} $E$ which is the expected value of the indicator when a center is performing according to the benchmark;
\item The {\em precision} with which the indicator is measured;
\item Control limits such that the chance of exceeding these limits for an in-control center is 5\%.
\end{enumerate} 

The funnel plot then shows each center with the precision on the x-axis and the ratio of observed to expected, $O/E$, the {\em indicator} on the y-axis. \cite{Spiegelhalter05} calls this $O/E$ the {\em target}; it is 1, i.e., the center is exactly on target if observed equals expected. The control limits mentioned above refer to the wish to formally test whether a center is performing according to the benchmark. In statistical terms, the objective is to test, for each center separately, the ``null hypothesis'' that that particular center's performance is equal to the benchmark. 

Of course, there will be ``case mix differences'' between centers in the sense that some centers will treat more high risk patients than others. This must be taken into consideration to ensure a fair comparison. One approach is to divide the patient population into more or less homogeneous groups, and then perform the benchmarking within those groups. Then, if a center shows particularly good or bad performance within a subgroup, the next step is to see if similar performance is observed in other subgroups. A clear disadvantage of this approach is that within each subgroup there are relatively few patients. As a result, there is limited statistical power to detect a performance difference with respect to the benchmark. Moreover, if we do the comparison in many subgroups, the probability of false positives increases. Of course, we could correct for multiple comparisons by using some method such as Bonferroni, but then the power is reduced even further. The lack of power to detect differences is a major problem because it will give centers a false assurance when in fact their performance may be sub-standard.

The alternative is to evaluate performance across all patients, but level the playing field by adjusting for patient characteristics using a multivariable regression and/or stratification. If a center shows particularly good or bad overall performance, the next step is to see if we can pinpoint any specific subgroup(s) where the center is over- or under-performing. To find such subgroups is a bonus, because it would provide a handle for possible steps toward improvement, but not strictly necessary. The success of this approach depends crucially on whether centers trust our ability to adjust for case mix differences. For this reason, we view the development of the case mix model as an ongoing process where the feedback of centers is taken into account to continually improve the model.

It is not difficult to adjust the funnel plot for case mix differences. We simply use the regression model to compute the expected value $E$ at each center, and compare it to the observed outcome. The funnel plot has become a popular tool in the assessment of healthcare quality; many clinicians have experience in interpreting funnel plots, and their use is accepted by statisticians because it incorporates the inherent uncertainty in the visualization. The goal of the present paper therefore is to extend the methodology of the funnel plot from binary outcomes to time-to-event outcomes. 

\subsection{Previous work}

The Center for International Blood and Marrow (CIBMTR) has performed mandatory benchmarking for many years now. For benchmarking survival outcomes they are using methodology based on pseudo-observations~\citep{Andersen03, Klein07}, as elaborated in Logan~\cite{Logan08}. The procedure was developed in order to benchmark survival at a fixed time point (one year in this case), without having to rely too heavily on the proportional hazards assumption. Briefly, the procedure starts by estimating, for each center, the one-year survival probabilities, by Kaplan-Meier. Then, for each subject in the data, pseudo-observations of one-year survival~\citep{Klein07} are calculated. These pseudo-observations are used in a generalized estimating equations (GEE) approach to incorporate the case mix. The authors used a Pearson-residuals based bootstrap procedure. In this procedure, bootstrapped values of the GEE model-based pseudo-observations are obtained, and these are aggregated over each center, leading to 95\% prediction intervals, assuming no center effects, of the expected proportion of survivors for each center. The statistical test for a center consists of comparing the Kaplan-Meier estimate of that center with the bootstrapped 95\% prediction interval, assuming no center effects. If the Kaplan-Meier survival estimate lies above the prediction interval the center is over-performing, if it lies below the prediction interval the center is under-performing, otherwise it is within target. It is important to note, however, that the prediction interval represents not only average performance, but also average follow-up. In a setting with differences in follow-up distributions between centers, if the follow-up distribution in the center of interest is very different from the average then the probability of hitting the interval may be very different from 95\%, also if the center is performing according to benchmark with respect to survival. Logan~\cite{Logan08} showed in simulation that the procedure performs adequately in a setting where there are no differences with respect to follow-up distribution between the participating centers. With the CIBMTR, for which the procedure was developed, this is indeed a reasonable assumption, since more than 95\% follow-up completeness is required in order to participate in the benchmarking process. For the EBMT setting however, currently this is not realistic, as we will see in Section~\ref{sub:benchfup}. In fact, the pseudo-observations approach~\cite{Logan08} is testing the null hypothesis that both performance and follow-up are the same as the benchmark. Our aim is to only test whether performance with respect to survival is different from the benchmark, and not also whether follow-up is different. Another downside of the pseudo-observations approach is that it does not easily allow for visualization of the center results in a funnel plot, which is the aim of the present work. 

Also \cite{hengelbrock2019evaluating} discuss benchmarking for time-to-event outcomes in the context of revisions of hip and knee arthroplasty implants and cardiac pacemakers. They propose two types of indicators, one of which is based on the center-specific survival function, the other based on a proportional hazards model with a multiplicative center-specific effect. They do not discuss the possibility of visualizing the center performance in a funnel plot. Their second proposal resembles our methods, but there are a number of important differences. We defer discussion of these differences of case mix correction, because some notation is necessary to appreciate them.

\vspace{0.5cm}
\noindent%
Here, we describe and discuss our methodology for the EBMT benchmarking project. In Section~\ref{sec:methods} we describe our proposed methods for constructing funnel plots for mortality, and we discuss practical issues, such as how to choose case mix variables, which centers to include. We also show how funnel plots for follow-up can be constructed along similar lines. In Section~\ref{sec:application} we illustrate our methodology using data from the EBMT benchmarking project. In Section~\ref{sec:simulation} we report on a simulation study centered around the EBMT setting. The paper concludes with a discussion in Section~\ref{sec:discussion}.

\section{Methods}
\label{sec:methods}

\subsection{Benchmarking survival outcomes}
\label{sub:benchsurv}

The context is center comparisons where the patient outcome of interest is $Y = {\bf{1}}\{ \tilde{T} \leq \tau \}$, for a fixed time-point $\tau$, for instance one year. Here $\tilde{T}$ is the event time of interest. The distinguishing feature of survival outcomes is the possibility of censoring, i.e., a subject has been followed until some censoring time $C$, and at that time the subject was known to be still alive. If there is no censoring before time $t$, then $Y$ is just a binary outcome and techniques for binary outcomes can be applied in a straightforward way, but the presence of censoring complicates matters in that we cannot observe $Y$ completely.

The data that are observed are realizations of $(T, D, X)$, where $T = \min(\tilde{T}, C)$ is the minimum of an event time $\tilde{T}$ and a censoring time $C$, $D$ is the status indicator (1 if the event occurred, 0 if the event time was censored), and $X$, an $n \times p$ matrix of baseline covariates, referred to as case mix. We have data of $n$ centers to be evaluated, indexed by $i=1,\ldots,n$, and $n_i$ patients within center $i$, indexed by $j=1,\ldots,n_i$. Thus, we observe realizations $(t_{ij}, d_{ij}, x_{ij})$ of $(T, D, X)$. The hazard of subject $j$ in center $i$ is given by $h_{ij}(t)$. 

Accounting for case mix $x_{ij}$, we will represent the benchmark by a proportional hazards model
\begin{equation}\label{eq:Cox}
    h_{ij}(t) = h_0(t) \exp(\beta^\top x_{ij}),
\end{equation}
with $h_0(t)$ a baseline hazard, and $\beta$ a $p$-vector of regression coefficients. In principle, the benchmark can be determined in any way that is deemed appropriate. We shall determine the benchmark by fitting a single Cox model to the complete data of all centers, disregarding the center identities, so that $h_0(t)$ and $\beta$ are the overall estimates obtained from fitting that model. Thus, the benchmark represents the average performance of all centers. We will ignore the uncertainty of the estimates of $h_0(t)$ and $\beta$, because it is negligible compared to the uncertainty arising from considering a single center, and also because the uncertainty in the benchmark is irrelevant. So, we will think of $h_0(t)$ and $\beta$ as being known, and use the known values in the computation of the cumulative hazards $H_{ij}(t) = \int_0^t h_0(s) \exp(\beta^\top x_{ij}) ds$, or, if the baseline hazard is a jump function, as obtained from a Cox model, $H_{ij}(t) = \sum_{0 < s \leq t} h_0(s) \exp(\beta^\top x_{ij})$, with $s$ the event time points, and $h_0(s)$ the Breslow estimate of the jumps in the cumulative baseline hazard. 

Define, for each patient $(i, j)$ (patient $j$ in center $i$) the counting process $N_{ij}(t)$, counting the number of events of patient $(i, j)$ before (and including) time $t$. The observed counting process is defined by $dN_{ij}^*(t) = dN_{ij}(t) Y_{ij}(t)$, where $Y_{ij}(t)$ is the at-risk indicator of patient $(i, j)$. Typically $Y_{ij}(t) = 1$ for $t \leq t_{ij}$, and is 0 afterwards, but the present set-up includes the possibility of left truncation, and even multiple (recurrent) events per subject. 

The censoring distribution will play a role in determining the expected number of events. Importantly, this distribution may differ between centers. Since we do not expect censoring in general to depend on case mix covariates, we define the center-specific hazard and the probability of being under follow-up at time $t$ by $h_{Ci}(t)$ and $G_i(t) = P(T_{ij} > t)$, respectively.

Set $\tau$ at the end of the follow-up period under consideration, and define, for center $i$, $O_i = \sum_{j=1}^{n_i} N_{ij}^*(\tau)$ as the total number of events observed during the assigned follow-up period over all subjects in center $i$. Under the null hypothesis of no difference in mortality with respect to the benchmark, after adjustment for case mix, $O_i$ will be a sum of independent Bernoulli random variables $N_{ij}^*(\tau)$, with expectations $p_{ij} = E N_{ij}^*(\tau)$. Here $p_{ij}$ is the  the probability of observing the event of interest for patient $(i,j)$ within the specified follow-up period. This probability can be expressed in terms of the hazard and survival functions of $T_{ij}$ and $C_{ij}$ as
\begin{equation}
\label{eq:pij}
    p_{ij} = P(T_{ij} < C_{ij}, T_{ij} < \tau) = \int_0^\tau h_{ij}(t) S_{ij}(t) G_i(t) dt.
\end{equation}
Note that here $h_{ij}$ and $S_{ij}$ depend on the patient characteristics $x_{ij}$ of patient $(i, j)$, while $G_i$ depends on the center. If the centers are not too small, the $G_i$ may be obtained by separately estimating the censoring distributions of each of the centers, for instance by reverse Kaplan-Meier.

Under the null hypothesis the expectation and variance of $O_i$ equal $E_i = \sum_{j=1}^{n_i} p_{ij}$ and $V_i = \sum_{j=1}^{n_i} p_{ij} (1 - p_{ij})$, respectively. Under the null, $O_i$ has the so-called Poisson-binomial distribution. The R package \texttt{poibin} implements that distribution, so exact $p$-values can in principle be calculated. Moreover, under reasonable assumptions the standardized random variable $(O_i - E_i)/\sqrt{V_i}$ approaches a standard normal distribution, as the sample size of the center $i$ tends to infinity. Informally, we write this as $O_i \sim N(E_i, V_i)$. This implies that $O_i/E_i \sim N(1, V_i/E_i^2)$. Still relying on asymptotic theory, a small improvement can be made by basing inference on $\sqrt{O_i}$ rather than on $O_i$~\citep[p.~163]{JohnsonKotzKemp}. In the EBMT setting and the simulations based on it, the normal approximation is adequate, so we have not pursued either exact $p$-values or normal approximations based on $\sqrt{O_i}$.

The intuitive explanation of ``Expected'' is the number of events expected in a center, based on the number of patients, their follow-up and their patient characteristics, when the center is performing according to the benchmark. The ratio $O_i/E_i$ is then the excess mortality; it may be interpreted as a standardized mortality ratio~\citep{berry1983analysis}. Based on the normal approximation, the asymptotic $\alpha$-level Wald test now becomes: reject $H_0$ for center $i$ if
\[
    \Bigl| \frac{O_i}{E_i} - 1 \Bigr| > \frac{z_{1-\alpha/2} \sqrt{V_i}}{E_i}.
\]
A funnel plot may then be created by plotting each center, with $E_i^2/V_i$ on the x-axis, ``Excess mortality'' $O_i/E_i$ on the y-axis, and the lines given by $f(x) = \pm z_{1-\alpha/2}/\sqrt{x}$.

The funnels are appropriate for each center in isolation, in the sense that the Type I error probability of incorrectly designating a center as either under- or overperforming is equal to $\alpha$. The probability that any of the centers falls outside the boundaries is of course much larger, due to multiple testing. Multiple testing adjusted funnels may also be added by adding the lines given by $g(x) = \pm z_{1-\alpha^\prime/2}/\sqrt{x}$, where $\alpha^\prime$ is a multiple testing adjusted nominal alpha level.

The quantity $E_i^2/V_i$ on the x-axis allows for the funnels (the function $f(x)$) to be plotted before plotting the data in the form of the center results. One could describe it as precision, but it is difficult to interpret. If there were no differences in case mix or censoring distributions between the centers, in other words if all $p_{ij}$'s would equal $p_0$, then $E_i^2/V_i$ would simplify to $\frac{(n_i p_0)^2}{n_i p_0 (1-p_0)} = n_i \frac{p_0}{1-p_0}$, which is just a multiple of the number of patients treated at center $i$. The common $p_0$ could simply be estimated as the mean over all patients and centers of the status indicator of death. If one then would put $E_i^2/V_i$ multiplied by $\frac{1-p_0}{p_0}$ on the x-axis, it would equal the sample size $n_i$ in case of no differences in case mix or censoring distributions between the centers. The quantity on the x-axis can be seen as an \emph{effective sample size}, adjusted for case mix and center follow-up. A funnel plot may then be created by plotting each center, with $E_i^2/V_i \cdot \frac{1-p_0}{p_0}$ on the x-axis, ``Excess mortality'' $O_i/E_i$ on the y-axis, and the lines given by $\tilde{f}(x) = \pm z_{1-\alpha/2} \sqrt{(1-p_0)/p_0}/\sqrt{x}$, possibly with multiple testing adjusted funnels given by $\tilde{g}(x) = \pm z_{1-\alpha^\prime/2} \sqrt{(1-p_0)/p_0}/\sqrt{x}$. The motivation for doing this is the familiarity of clinicians with ``standard'' funnel plots, for instance in the binary and normal case, that have sample size on the x-axis.

\subsection{Benchmarking follow-up}
\label{sub:benchfup}

Adequate data quality is essential for reliable benchmarking. This includes completeness of the registration of those risk factors determined to be used in the case mix models, and in the context of benchmarking survival outcomes also completeness of follow-up. Informative registration of deaths, where deaths are reported in a center, but follow-up of those alive is lagging, will result in possibly serious bias, to the disadvantage of the center. Measures to quantify follow-up completeness exist~\citep{clark2002quantification}, but completeness for follow-up relative to the centers being benchmarked for mortality may also be visualized in a funnel plot, by reversing the role of event and censoring, as in the reverse Kaplan-Meier.

Define, for each patient $(i, j)$ (patient $j$ in center $i$) the counting process $N_{ij}(t)$, counting the number of events of patient $(i, j)$ before (and including) time $t$. The observed counting process is defined by $dN_{ij}^*(t) = dN_{ij}(t) Y_{ij}(t)$, where $Y_{ij}(t)$ is the at-risk indicator of patient $(i, j)$. Typically $Y_{ij}(t) = 1$ for $t \leq t_{ij}$, and is 0 afterwards, but the present set-up includes the possibility of left truncation, and even multiple (recurrent) events per subject. 

Extending the notation of Section~\ref{sub:benchsurv}, we define for each patient $(i, j)$ the counting process $\tilde{N}_{ij}(t)$, counting the number of losses of follow-up (either 0 or 1) of patient $(i, j)$ before (and including) time $t$, and use $\tilde{O}_i = \sum_{j=1}^{n_i} \tilde{N}_{ij}(\tau)$ as the total number of losses to follow-up during the assigned follow-up period over all subjects in center $i$. Under the null hypothesis of no differences in follow-up distributions between the centers, $\tilde{O}_i$ will again be a sum of independent Bernoulli random variables $\tilde{N}_{ij}(\tau)$, with expectation $\tilde{p}_{ij} = E \tilde{N}_{ij}(\tau)$, the probability of observing the loss to follow-up within the specified follow-up period. This probability can be expressed in terms of the hazard and survival functions of $T_{ij}$ and $C_{ij}$ as
\begin{equation}
\label{eq:tildepij}
    \tilde{p}_{ij} = P(C_{ij} < T_{ij}, C_{ij} < \tau) = \int_0^\tau h_{C}(t) G(t) S_{ij}(t) dt.
\end{equation}
Although Equations~\eqref{eq:pij} and~\eqref{eq:tildepij} look almost identical, there are subtle differences, because of the fact that a different null hypothesis is being tested. In particular, since we are no longer working under the null hypothesis of no differences in mortality rates between the centers, in Equation~\eqref{eq:tildepij} $S_{ij}(t)$ depends not only on the patient characteristics $x_{ij}$ of patient $(i, j)$, but also on the center. That means that for calculating $\tilde{p}_{ij}$ we have to fit a Cox model with the case mix variables and either fixed or random center effects, or by stratifying on center. Since we work under the null hypothesis of no differences in follow-up distributions between the centers, we have a single hazard $h_C(t)$ and probability of being under follow-up $G(t)$ which can be estimated by reverse Kaplan-Meier, using the pooled data. After having calculated the $\tilde{p}_{ij}$'s the funnel plot proceeds along the same lines as the funnel plots for mortality.

\subsection{Practical issues}
\label{sub:practical}

Several choices need to be made when implementing benchmarking for survival outcomes. Most of these are actually not specific to survival outcomes. Each choice is discussed in general first, and then we report on how we dealt with them in the EBMT benchmarking project.

The first choice to be made is which variables are to be included in the case mix correction model. Key is that the comparison of the center with the benchmark is fair and is not confounded by differences in patient characteristics between the centers. The general rules that epidemiologists use to control for confounding~\citep{rothman2008modern, vanderweele2019principles} dictate that all confounding factors should be included in the case mix model, and that choices whether or not to include a patient characteristic should be made primarily based on subject-matter knowledge, and not based on $p$-values or predictive accuracy. Importantly, factors that are on the causal pathway between centers and outcome should {\em not} be included in the case mix correction model. In the context of benchmarking, such factors would include variables that can be influenced by the center, such as decisions whether or not to treat a subgroup of patients in a certain way. For the EBMT benchmarking project, a Clinical Outcomes Group was set up to decide on the case mix variables to be included. The items used by the CIBMTR were adopted, subject to availability in the EBMT registry. The benchmark model for allogeneic transplantations included an adaptation of the Disease Risk Index (DRI)~\cite{armand2014validation}, calculated from the diagnosis/disease status info, in order to include a risk factor based only on disease type and status at the time of transplantation, and including cytogenetics for AML/MDS. Also, the following variables were included: as recipient variables age, sex, coexisting disease (HCT-specific comorbidity index, HCT-CI), cytomegalovirus (CMV) serological status, Karnofsky/Lansky performance status at transplant, prior autologous transplant, donor variables age, patient-donor sex match, donor type (matched sibling donor vs.~matched related vs.~mismatched related vs.~unrelated donor), and general variables first complete remission (CR1) vs CR$>$1 vs not in CR for AML and ALL, all others combined as not AML/ALL (interval between diagnosis and transplant and a dummy for AML/ALL both included as covariates), and year of transplant. For autologous HSCT, only recipient age, sex, Karnofsky/Lansky, year of transplant and DRI were included. Trust of the stakeholders in the fairness of the funnel plot is of key importance, so decision on whether or not to include further variables in the case mix correction model are carefully considered by the Clinical Outcomes group.

The second issue to be discussed is missing values. We argue that missing case mix data should be dealt with differently when fitting the case mix models and when performing the actual benchmarking. For fitting the benchmark models, multiple imputation should be used to avoid any bias due to missingness at random. For the EBMT benchmarking project we used MICE (multiple imputation by chained equations). When doing the actual benchmarking in the EBMT project, we imputed missing case mix variables by their median value among all patients within the EBMT with observed favorable outcome (for benchmarking one-year mortality this means patients that survived one year after HSCT). This will make the patient appear ``relatively healthy'', decrease the expected number of events in the center, and lead to an unfavorable observed over expected ratio for the centre. The idea is that this should encourage centres to strive for complete registration of case mix variables.

The third issue is which centers to include in the benchmarking. In fact there are two choices to be made. The first is which centers are to be used for the case mix model. For this step only centers with reliable, complete data, should be used, again because a fair comparison between the center and the benchmark is crucial. The second choice is which centers are to be benchmarked. In principle all centers could be included in this step, but a minimum volume could be imposed if the tests to be used rely on asymptotic theory. An alternative would be~\cite{hengelbrock2019evaluating}, to use exact $p$-values, in which case all centers with adequate data could be included. For the EBMT benchmarking project, precise inclusion criteria for centers are reported in a position paper~\citep{snowden2020benchmarking}; selections were made separately for allogeneic and autologous transplants and included a minimum of 10 allogeneic and 5 autologous transplants on average per year during the 2013--2016 period, and a minimum of 80\% of the transplants to be reported in the EBMT Activity Survey. 

The fourth issue is choosing the population of patients. It makes sense to leave out certain rare subgroups of patients for which the comparison of the center against the benchmark is not appropriate; this is primarily a decision to be made by the clinical experts, based on subject-matter knowledge. In the EBMT benchmarking project, the Clinical Outcomes Group decided to include only first autologous and first allogeneic HSCT (including those preceded by an autologous transplant). In addition, autologous HSCT for solid tumors indications were excluded. For autologous HSCT, only transplants for adults with haematological cancers were included.

Finally, should socio-economic factors be included in the case mix correction? This is a difficult issue, and the answer probably depends on the context. Based on our discussion of factors to be included in the case mix model, they should, because they are confounders. Nevertheless, for the EBMT benchmarking project we decided not to pursue this, because (1) socio-economic factors are very hard to adequately capture and (2) we want to {\em show} how for instance under-funded centers are struggling; our aim is not to know how the centers would perform in case of equal funding.

\section{Application}
\label{sec:application}

The Joint Accreditation Committee ISCT-Europe \& EBMT (JACIE) is Europe's only official accreditation body in the field of haematopoietic stem cell transplantation (HSCT) and cellular therapy. The EBMT benchmarking project was initialized in 2018, when the department of Biomedical Data Sciences of the Leiden University Medical Center was appointed by JACIE to lead the statistical analysis underlying annual cycles of reports to be sent to each of the transplant centers performing autologous HSCT's in adults or allogeneic HSCT's. The benchmarking methodology was to incorporate a series of risk factors (case mix variables) to be integrated into the statistical models to allow for a fair comparison of centers related to different patient population characteristics. The output was to be a risk-adjusted comparison of each center with the internal benchmark, set by the average across participating EBMT centers. The selection of case mix variables for the ``first phase'' that we report on here was based on an appraisal of the available data and subsequent consensus across a ``clinical outcomes'' group, consisting of senior HSCT clinicians, registry managers, EBMT (including JACIE) staff and biostatisticians from LUMC, EBMT Patient Advocacy Committee and national societies.

For decisions on how to deal with missing case mix data and which patients and centers to include we refer to Section~\ref{sub:practical}. We report here on allogeneic transplants only; for results on autologous transplants we again refer to the position paper~\cite{snowden2020benchmarking}. During the 4-year period 2013--2016, a total of 288 centers, with a total of 49,612 patients, contributed to the benchmarking project for allogeneic transplants.

\begin{figure}[ht]
\centering
  \includegraphics[width=0.9\textwidth]{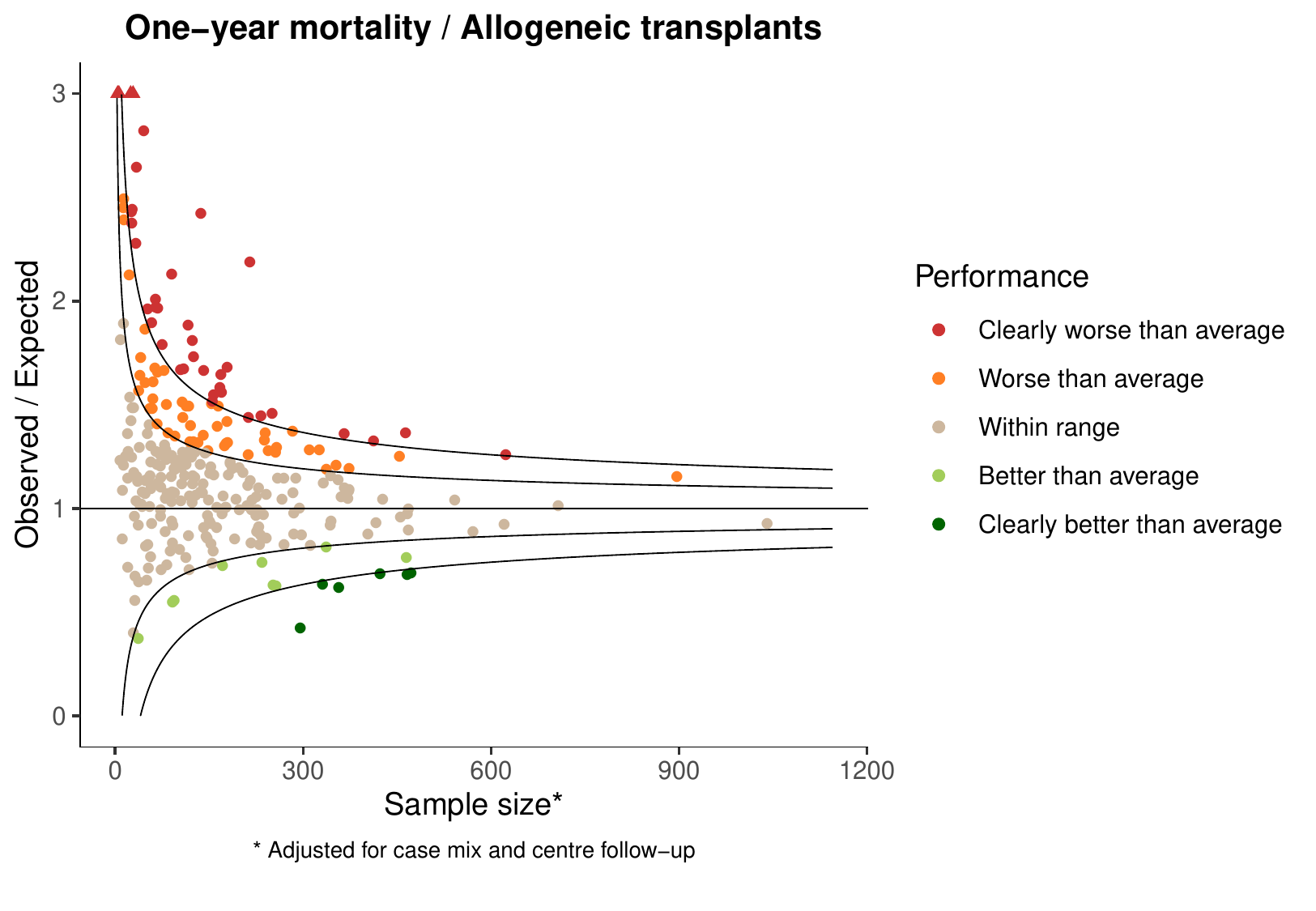}\\
  \caption{Observed / expected representation of funnel plot of death within one-year}\label{fig:funnelOS1}
\end{figure}

Figure~\ref{fig:funnelOS1} shows the funnel plot for one-year mortality for allogeneic stem cell transplantations in the EBMT.
The sample size reported along the x-axis is the effective sample size, detailed in Section~\ref{sub:benchsurv}, calculated over the four-year period. The majority of centers (184, 63.9\%) fall within the range set by requiring that under the null hypothesis 95\% of centers falls within the range. A total of 89 centers (30.9\%) performs worse than average, of which 38 centers (13.2\% of total) perform clearly worse than average. The number of centers that perform better than average is 15 (5.2\%), of which 6 (2.1\% of total) perform clearly better than average. There is clearly more variability in the center's performance than expected under the global null hypothesis.

\begin{figure}[ht]
\centering
  \includegraphics[width=0.9\textwidth]{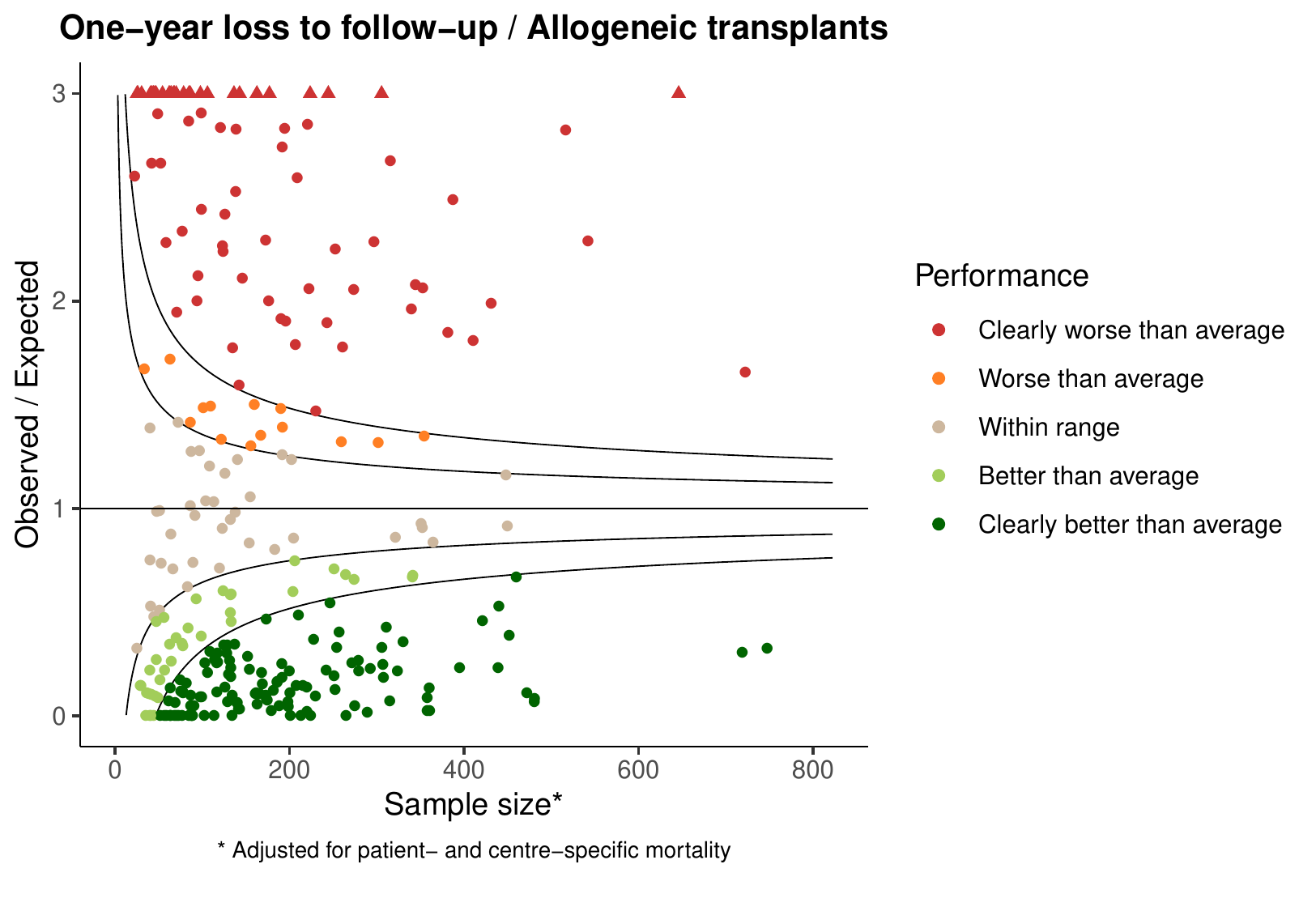}\\
  \caption{Observed / expected representation of funnel plot of loss to follow-up within one-year}\label{fig:funnelfup1}
\end{figure}

The higher variability than expected under the global null is even more extreme when looking at one-year loss to follow-up. Figure~\ref{fig:funnelfup1} shows the funnel plot for one-year loss to follow-up for the allogeneic stem cell transplantations in the EBMT.
Here only a minority of centers (39, 13.5\%) falls within range. A total of 91 centers (31.6\%) performs worse than average, of which 77 centers (26.7\% of total) perform clearly worse than average. The number of centers that perform better than average is 158 (54.9\%), of which 121 (42.0\% of total) perform clearly better than average. Clearly, effort is needed to improve adequate collection of follow-up data for a substantial number of centers. The funnel plot for follow-up in Figure~\ref{fig:funnelfup1} shows performance of each center compared to the EBMT average as benchmark. In fact, we would like the completeness to be better than the current average. If we had benchmarked against, say, 95\% completeness, which is the requirement for inclusion for benchmarking in the CIBMTR, only a very small minority of centers would have met the standard.

\section{Simulation study}
\label{sec:simulation}

\subsection{Set-up}
\label{sub:setup}

The set-up of the simulation study is based on the data describing allogeneic transplants of the EBMT application in Section~\ref{sec:application}. The base scenario used 300 centers, sample size per center was generated from a negative binomial distribution with mean and standard deviation as estimated from the allogeneic EBMT data, namely 200 and 150, respectively. Time was measured in months (since HSCT). Censoring distributions were generated from separate Weibull distributions per center; log shape and log rate were generated from a multivariate normal distribution with mean 0.4 and -4.8, respectively, and standard deviation 0.24 and 1.72, respectively, with correlation -0.87. These numbers were obtained by fitting separate Weibull distributions to the censoring distributions of the centers. Modeled after the linear predictor of the Cox case mix model for one-year mortality, a single covariate $x$ was generated, with regression coefficient equal to 1. The distribution of $x$ was taken to be normal, with between-centers variance and within-centers variance equal to the estimated between-centers variance and within-centers variance of the linear predictor in the case mix model for the allogeneic EBMT data, namely 0.056 and 0.224, respectively. The base scenario had the same Weibull baseline distribution for all the centers, with shape 0.94 and rate 0.032, as obtained from the allogeneic EBMT data after fitting a Weibull regression to the data of all centers combined with the linear predictor as sole covariate. Since each replication in the base scenario already contains data on 300 centers and 60,000 patients on average, we used 50 replications. The Monte Carlo standard error for the coverage probability under the basic scenario was $0.56\%$.

The base scenario was altered in a number of ways to study different aspects of our approach. The effect of sample size was assessed first by changing the number of centers to 30, keeping the distribution of the number of patients per centers the same (``Fewer centers''), then by changing mean and variance of the number of patients per center to 20 and 15, respectively, keeping the number of centers the same (``Fewer patients''). For these two scenarios, 500 replications were used. Average power was assessed by multiplying the baseline Weibull rates for mortality by log-normal frailty terms (one independent realization for each center), with variance of the log-frailty equal to 0.15 (which is the variance of the log-frailty after fitting a log-normal frailty model to the EBMT data, with inclusion of the linear predictor of the case mix model) and 0.3; these two scenarios are referred to as ``Small frailty'' and ``Large frailty'', respectively. The base scenario used different censoring distributions across centers; a simpler scenario was included (``Base same fup'') for which the censoring distribution was the same across centers, namely Weibull with shape and rate parameters equal to $\exp(0.4)$ and $\exp(0.8)$, respectively. Finally, the effect of moderate deviations from proportional hazards was studied by multiplying the baseline Weibull rates for mortality by log-normal frailty terms (one independent realization for each center), with variance of the log-frailty equal to 0.15, and changing the Weibull shapes in such a way that the 12 months baseline survival probabilities were the same for all centers. Again, 50 replications were used. For each of the alternative scenarios, all other parameters were kept the same as in the base scenario. 

In each of the replications, a Cox regression with the linear predictor was fitted to the overall data, following the methods outlined in Section~\ref{sub:benchsurv}, results were aggregated per center, recording observed and expected deaths $O_i$ and $E_i$ within $\tau=12$ months, as well as the variance under the null hypothesis $V_i$. From these, $Z_i = (O_i - E_i)/\sqrt{V_i}$ were calculated, and the number of centers for which $Z_i$ was less than $-z_{0.975}$ (``Over'' for over-performing), more than $z_{0.975}$ (``Under''), and between $-z_{0.975}$ and $z_{0.975}$ (``Target'') were recorded. The pseudo-observations approach~\cite{Logan08} was assessed by estimating for each center the one-year mortality probabilities by Kaplan-Meier. The Pearson-residuals based bootstrap procedure, proposed in~\cite{Logan08}, was used (with 1000 bootstrap replications) to obtain 95\% prediction intervals, assuming no center effects, of the expected number of deaths within one year for each center. We recorded the number of centers for which the Kaplan-Meier estimate of one-year death probability was below (``Over'' for over-performance), above (``Under''), and within (``Target'') the 95\% prediction interval.

\subsection{Results}
\label{sub:results}

Table~\ref{tab:simresults} shows the results of the simulation study. The columns under ``Funnel'' show the results for our proposed methodology. Mean and standard deviation of the Z-scores are close to the target values of 0 and 1, for the first five scenarios, where there are no differences in adjusted performances between the centers. In the base scenario and the scenarios with fewer centers and fewer patients (for settings see previous subsection) there is a conservative tendency, with over-performance being detected in two percent of centers. This conservative behavior seems to be primarily a small-sample issue, since it is more serious for the settings with fewer centers and fewer patients. The pseudo-observations approach~\cite{Logan08} suffers from more serious anti-conservative behavior, which is due to the fact that differences across centers in follow-up distribution is not accounted for. We will return to this issue at the end of this section. In the ``Base same fup'' setting, where the follow-up distribution was taken to be the same for all centers, the method was performing adequately. For the EBMT setting, unfortunately, this is not a realistic scenario. Our proposed methodology appears to be robust to moderate deviations from the proportional hazards assumption, as shown in row ``Non-PH''. The bottom two rows show the power (averaged over centers) to detect outlying centers (both under- and overperforming), in case the variance of the log-frailty equals 0.15 (as in the EBMT allogeneic data, ``Small frailty'') and 0.30 (twice that of the EBMT allogeneic data, ``Large frailty''). Both our approach and the pseudo-observations approach identify approximately half of the centers as under- or over-performing. The pseudo-observations approach seemingly has larger power, but this is not to be taken as evidence of the pseudo-observations approach being superior, since its type-I error was too high under the null.

\begin{table}[ht]
    \centering
    \begin{tabular}{l|rrrrr|rrr}
  \hline\hline
  & \multicolumn{5}{c}{Funnel} & \multicolumn{3}{c}{Pseudo} \\
  & \multicolumn{2}{c}{Z-scores} & \multicolumn{3}{c}{Percentages} & \multicolumn{3}{c}{Percentages} \\
  & \multicolumn{1}{c}{Mean} & \multicolumn{1}{c}{SD} & \multicolumn{1}{c}{Under} & \multicolumn{1}{c}{Target} & \multicolumn{1}{c}{Over} & \multicolumn{1}{c}{Under} & \multicolumn{1}{c}{Target} & \multicolumn{1}{c}{Over} \\
  \hline
  Base & -0.001 & 0.982 & 2.6 & 95.4 & 2.0 & 4.5 & 92.2 & 3.3 \\
  Base same fup & 0.002 & 0.989 & 2.5 & 95.4 & 2.1 & 2.7 & 94.9 & 2.4 \\
  Fewer centers & 0.006 & 0.966 & 2.4 & 96.0 & 1.6 & 4.0 & 92.6 & 3.4 \\
  Fewer patients & -0.001 & 0.985 & 3.0 & 95.8 & 1.2 & 3.9 & 92.9 & 3.2 \\
  Non-PH & 0.003 & 1.018 & 2.8 & 94.7 & 2.5 & 4.3 & 92.3 & 3.4 \\
  \hline
  Small frailty & 0.006 & 2.836 & 20.9 & 56.6 & 22.5 & 24.5 & 55.6 & 20.0 \\
  Large frailty & 0.003 & 3.814 & 25.1 & 44.7 & 30.1 & 31.9 & 43.4 & 24.7 \\
  \hline\hline
\end{tabular}
    \caption{Results of simulation study}
    \label{tab:simresults}
\end{table}

\begin{figure}[ht]
\centering
  \includegraphics[width=0.6\textwidth]{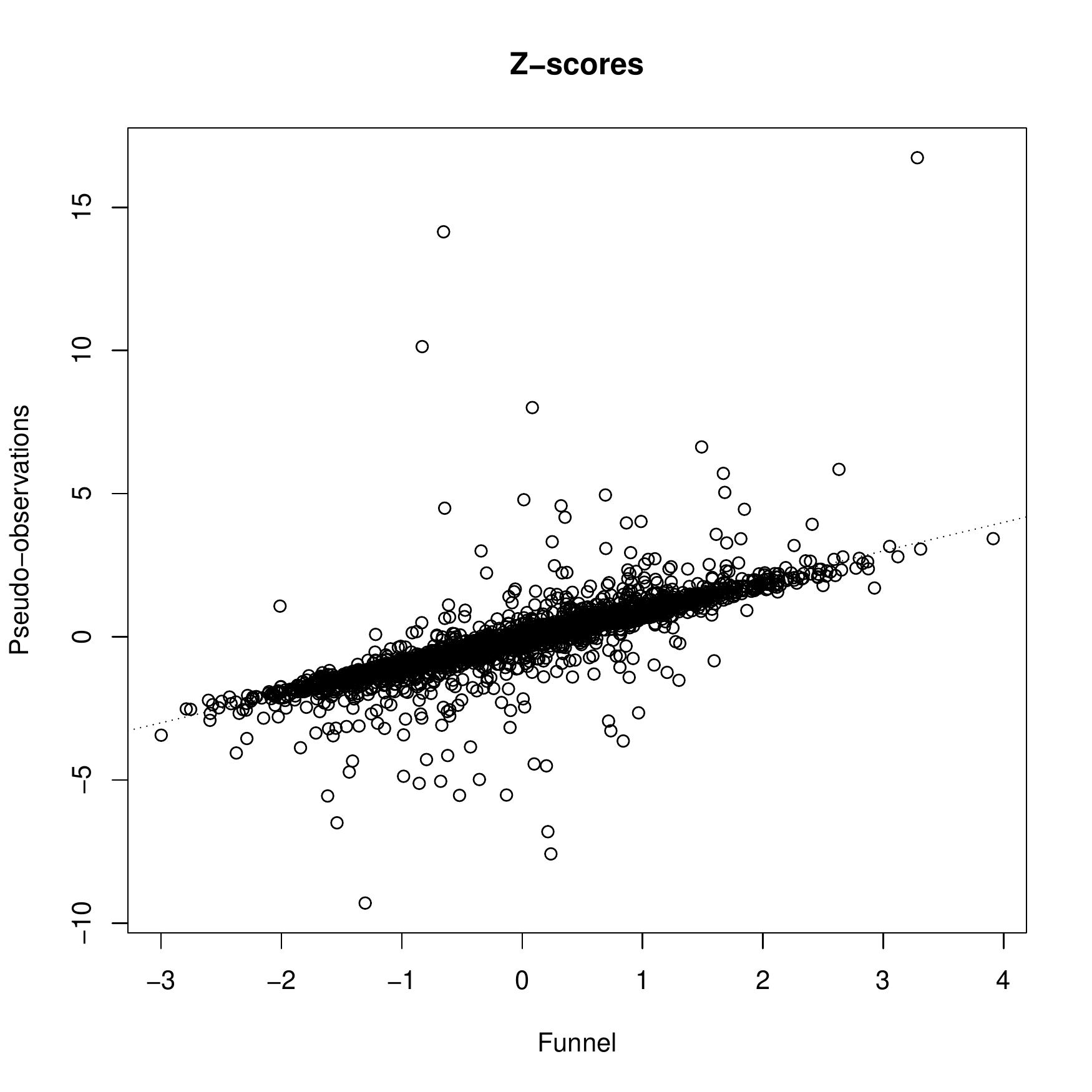}\\
  \caption{Z-scores of the funnel plot versus the pseudo-observations approaches}
  \label{fig:Zscores}
\end{figure}

\begin{figure}[ht]
\centering
  \includegraphics[width=0.49\textwidth]{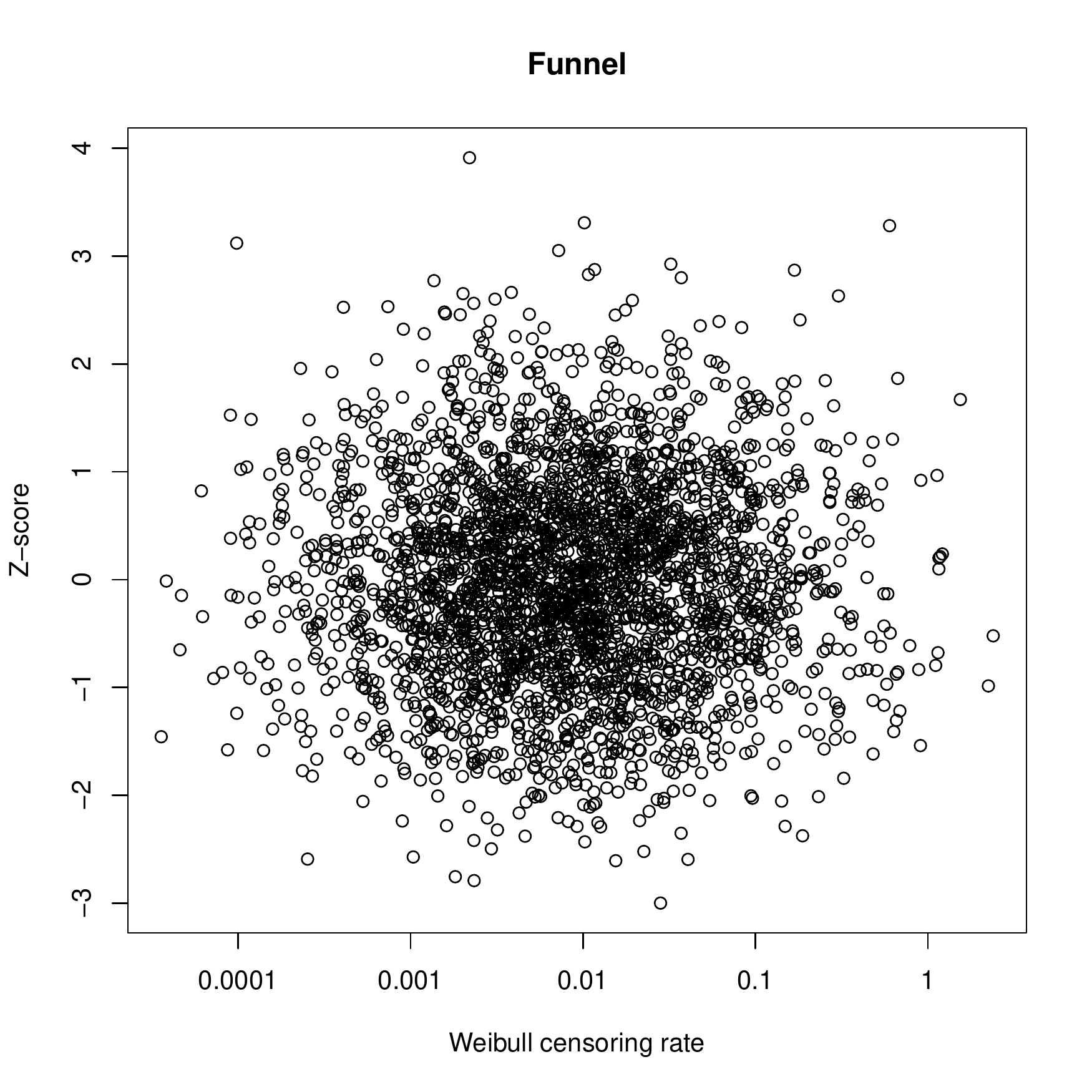}
  \includegraphics[width=0.49\textwidth]{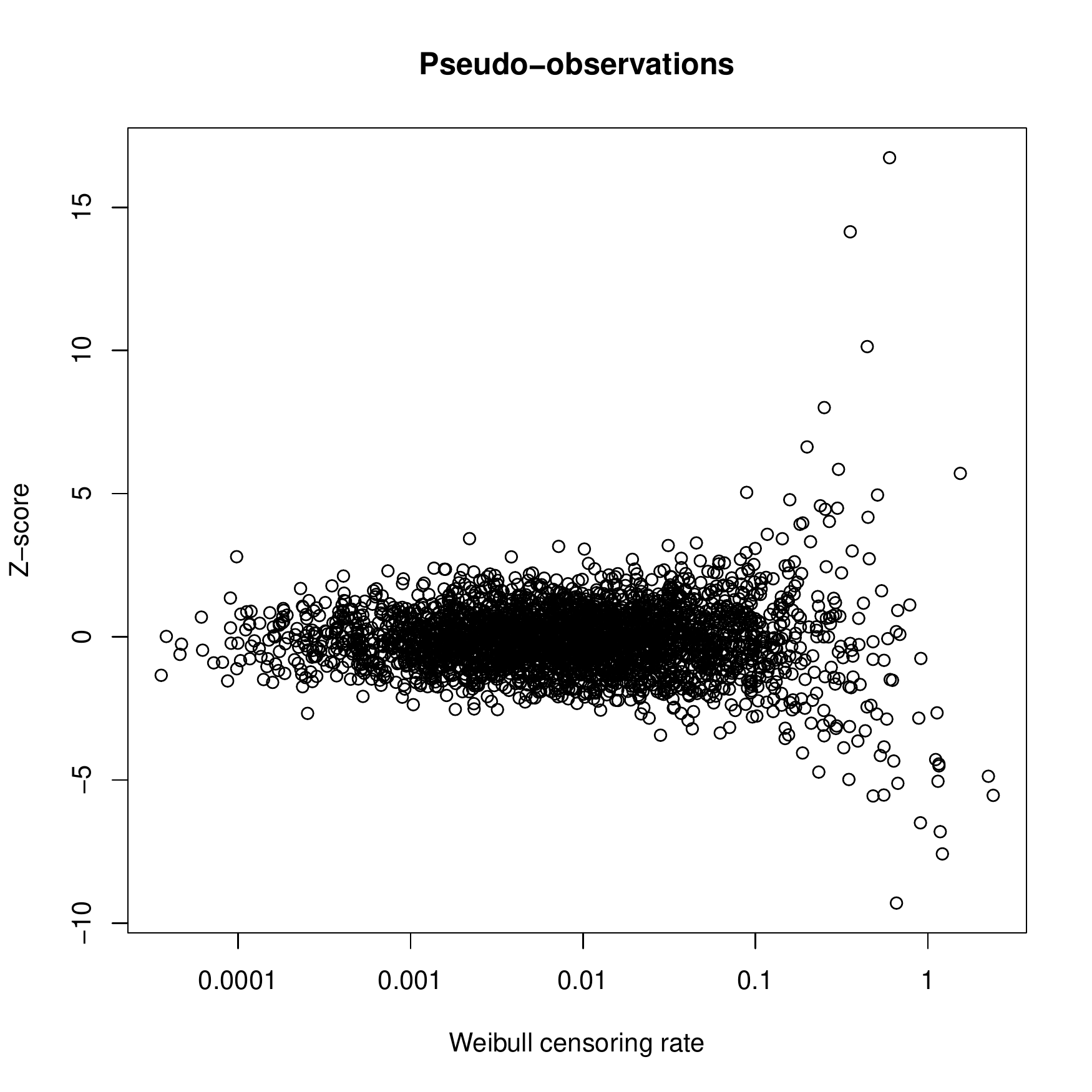}\\
  \caption{Z-scores of the funnel plot and the pseudo-observations approach versus the Weibull censoring rates}
  \label{fig:Zscores_b}
\end{figure}

It is worthwhile trying to understand the issue of differences in follow-up distributions and the pseudo-observations approach. In that approach the ``observed'', the Kaplan-Meier estimate at the time point of interest, is compared with a prediction interval, based on bootstrapping residuals from a GEE model using pseudo-observations. The width of this prediction interval is partly determined by the length of follow-up; for centers with long follow-up the prediction interval will be narrower than for centers with short follow-up. We have repeated the simulation of the base scenario with a smaller number of replications (ten, each with 300 centers). For each of the replications, based on the same data, we recorded the Weibull shape and rate parameters of the censoring distributions, as well as the Z-scores obtained by the funnel plot procedure. The pseudo-value approach does not directly yield Z-scores, but these were calculated from the 2.5\% and 97.5\% quantiles of the prediction intervals, assuming a normal distribution of the distribution of ``observed'' if the center is performing according to benchmark (for instance, the 2.5\% and 97.5\% quantiles of the prediction interval would yield Z-scores of -1.96 and 1.96 and the middle of the interval a Z-score of 0). Figure~\ref{fig:Zscores} shows a scatterplot of the Z-scores of the funnel plot (called ``Funnel'') and that of the pseudo-observations approaches. Agreement between the two Z-scores is generally very high, with the exception of a number of centers where the pseudo-observations approach gives a very high Z-score.
Figure~\ref{fig:Zscores_b} shows scatterplots of the rate parameters of the censoring distributions against the Z-scores of the funnel plot (a) and that of the pseudo-observations approach (b). It can be seen that the variability of the funnel plot Z-scores is independent of the censoring distribution rates, while the variability of the pseudo-observations Z-scores is not. The variability of the pseudo-observations Z-scores is comparable to that of the funnel plot Z-scores in the middle range of the censoring rates; in the lower range of the censoring rates, however, the variability is much smaller, while it is much too large in the upper range of the censoring rates. This leads to a too high proportion of false rejections of the null hypothesis of the center performing according to benchmark when the censoring rate is high.

\section{Discussion}
\label{sec:discussion}

In this paper we have proposed methodology for constructing funnel plots for survival data. Simulation studies show that the method has adequate type I error control under the setting used in the EBMT, which includes differences in follow-distributions across centers, and in several deviations from this setting, including smaller number of centers, smaller center size, and deviations from the proportional hazards assumption. The funnel plot is an attractive tool for the assessment of center performance with respect to time-to-event outcomes, because of its familiarity in other healthcare quality assessment settings, and because it allows visualization of both effect size and statistical uncertainty. By reversing the role of event and censoring indicator, similar to the reverse Kaplan-Meier for estimating the follow-up distribution, the same ideas for constructing funnel plots for mortality can also be used to construct funnel plots for follow-up.

Needless to say: the proposed procedure stands or falls with the availability of high quality data. This goes for completeness and reliability of the case mix variables and of follow-up, hence our suggestion to benchmark follow-up prior to benchmarking mortality. One issue related to completeness of follow-up is preferential reporting of events; centers might be inclined to prioritize providing data in the registry about deaths, without making sure that comparable attention is paid to providing data about follow-up without events. This will result in a bias which is unfavorable for the center, because compared to the full information the ``observed'' number of events remains unchanged, but the ``expected'' is reduced due to shorter follow-up, leading to a higher observed over expected ratio. Of course, if deaths remain unreported, bias is introduced in the other direction. Note that this potential bias is not really specific to our proposed methodology.

A limitation of our approach is that we are comparing at some arbitrary time point, in our application one year. On the other hand, pre-defining such a time horizon is probably wise, since otherwise differences between centers with regard to follow-up will play a bigger role. Also, limiting the follow-up to a fixed time point may make the procedure more robust against violations of the proportional hazards assumption. The simulation study of Section~\ref{sec:simulation} showed that moderate deviations from the proportional hazards assumption are not harmful in the EBMT setting, but it should be acknowledged that this is in a setting where the follow-up is quite short anyway (one year). More study is needed to evaluate our methods in a setting with long follow-up and more severe violations of the proportional hazards assumption. When administrative censoring is applied at the time point of interest for the benchmarking, a procedure sometimes called ``stopped Cox'', it is known that Cox models can still be used to obtain approximately valid predictions of survival at that same time point, even under violations of proportional hazards~\citep{vHP12, stoppedcox}. It is unclear how including the follow-up distribution in addition to the formula leading to prediction in the stopped Cox context (Equation~\eqref{eq:pij} without the $G_i(t)$ term) would work out.

It would be of interest to extend our methods to competing risks. In the context of HSCT, the two competing risks relapse and non-relapse mortality are of central interest. In principle, this should be feasible, by adapting Equation~\eqref{eq:pij} using the cause-specific hazard of the cause of interest and the event-free survival function.

The method of~\cite{hengelbrock2019evaluating} also works with a ratio $O/E$, as does our approach, and could therefore in principle also be used as input for a funnel plot. Their expected $E$ is defined differently from ours; it is defined as the sum over individuals in the center of the cumulative patient-specific hazards evaluated at the observed time points, assuming no center differences. Its {\em expectation} equals our definition of $E$, but the term itself is random, through the use of the observed time points. This additional randomness makes it more difficult to rely on the normal approximation. The authors use a likelihood ratio test (twice the log-likelihood evaluated at $O/E$ minus the log-likelihood evaluated in 1, which is then compared with a $\chi^2$ distribution with one degree of freedom) to test whether the center is performing according to the benchmark. 

The ultimate goal of healthcare quality assessment is improvement of patient care. We have provided a tool for centers to get more insight into their own performance, allowing them to gauge how they are doing in comparison with their peers, after correcting for possible differences in case mix. The EBMT benchmarking project is now entering its ``second phase'', after having sent out initial reports and incorporating feedback received from the participating centers. Trust and transparency of any benchmarking enterprise is essential, both in the procedure and in the statistical models used. We must be modest in what we claim; no case mix correction model will be perfect, and we should be aware that failure to account for important variables could possibly result in false positive results for a center. On the other hand, even an imperfect case mix correction model is to be preferred over a crude comparison. 

\subsection*{Acknowledgement}

The authors would like to thank Per Kragh Andersen for helpful discussions about incorporating the censoring distribution.

\subsection*{Availability of data}

The EBMT data set used for illustration in the present manuscript
cannot be made available for reasons of confidentiality.

\bibliographystyle{WileyNJD-AMA}
\bibliography{benchsurvbib}

\end{document}